# REDUCTION OF CONSTRAINT SYSTEMS

**Samy Ait-Aoudia, Roland Jegou, Dominique Michelucci**
Ecole des Mines de Saint Etienne
158 cours Fauriel
42023 ST ETIENNE cedex 2
FRANCE
ait@emse.fr, jegou@emse.fr, micheluc@emse.fr

## ABSTRACT

Geometric modeling by constraints leads to large systems of algebraic equations. This paper studies bipartite graphs underlaid by systems of equations. It shows how these graphs make possible to polynomially decompose these systems into well constrained, over-, and under-constrained subsystems. This paper also gives an efficient method to decompose well constrained systems into irreducible ones. These decompositions greatly speed up the resolution in case of reducible systems. They also allow debugging systems of constraints.

**Key Words:** geometric modeling, constraints, bipartite graphs, matching, maximum matching, perfect matching.

## 1. INTRODUCTION

Geometric modeling by constraints is an interesting approach in CAD. Typically, in 2D, geometric modeling by constraints specifies geometrical objects such as points, lines, circles, conics by a set of constraints : distances between points, points and lines, parallel lines, angles between lines, incidence relations between points and lines, points and circles, tangency relations between lines and circles or between circles, and so on. In 3D, geometric modeling by constraints must take into account new objects like planes, quadrics, and new constraints such as dihedral angles.

Geometric modeling yields large systems of algebraic equations (linear or not linear). Many programming styles or languages have been investigated : imperative, object oriented, rules driven. Many resolution methods have been investigated : geometric, numerical, symbolic... Geometric methods can be very efficient but are only applicable to particular kinds of problems : see [Owen91]. Numerical methods (Newton iteration, gaussian elimination, matrix inversion and so on) are $O(N^3)$ or worse. Symbolic methods (Grobner bases, elimination with resultants) are typically exponential in time and space. Report to the survey in [Verroust90] and [Roller et al.89] for details on these different approaches. Anyway, all general methods are time-consuming on large systems, so any reduction method is interesting.

This paper considers the natural bipartite graph associated with systems of equations and gives some structural properties of this graph, which can be used to simplify resolution. This bipartite graph has one vertex per equation, one vertex per unknown, and an edge between an unknown $x$ and an equation $y$ iff $x$ appears in equation $y$. This type of graphs has been already used, for example by Serrano in [Serrano91]. By convention, equation vertices are elements of $Y$, unknowns are elements of $X$, and in all figures, equation vertices are drawn above unknown vertices.

In the general case, this graph is sufficient enough to decompose the system of equations in three parts : well constrained, over-constrained and under-constrained subsystems. Some of these parts can be empty. This decomposition, which always exists and is unique, is due to Dulmage and Mendelsohn.

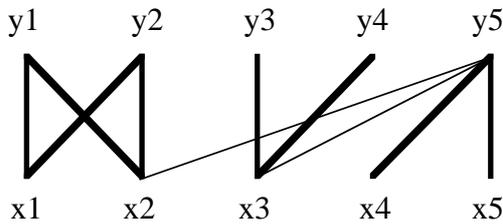

**Figure 1.**

In Figure 1, $\{y_1, y_2, x_1, x_2\}$ is a well constrained subsystem. $\{y_3, y_4, x_3\}$ is an over-constrained subsystem. $\{y_5, x_3, x_5\}$ is an under-constrained subsystem.

Let us first give some intuitive definitions. An over-constrained system has more equations than unknowns; its equations are either redundant, or contradictory and thus yield no solution. An under-constrained system has more unknowns than equations and has generally an infinite and not enumerable set of solutions. In a well constrained system, the number of equations is equal to the number of unknowns, the system contains no over-constrained subsystem, and it has a finite number of solutions. The associated graph is said to be well constrained (respectively under-, over-) when its system is well constrained (respectively under-, over-).

In a mathematical way, a system is well constrained iff its associated bipartite graph satisfies the König-Hall relation : for any subset $Y'$ of the set $Y$ of equations, we have $|\Gamma(Y')| \geq |Y'|$, where $\Gamma(Y')$ is the set of all neighbours of $Y'$, ie the set of unknowns appearing in $Y'$, and $|\ |$ is the cardinality. This condition is equivalent to the existence of a perfect matching, see later in the paper.

Let us introduce a second decomposition.

A well constrained graph $G$ is said to be irreducible iff for any proper subset $Y'$, we have $|\Gamma(Y')| > |Y'|$. Let us call a down-subgraph any subgraph of $G$ induced by $Y'$ and $\Gamma(Y')$, where $Y'$ is a subset of $Y$. It is easy to show that any well constrained graph is either irreducible, or contains an irreducible down-subgraph $I$. Moreover $G - I$ is still well constrained, or empty. So the resolution of the system associated with $G$ is reduced to the successive resolution of the subsystems associated with the irreducible subgraphs of $G$.

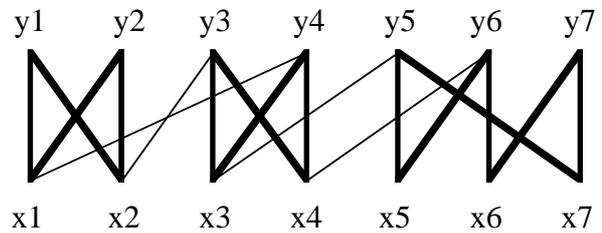

**Figure 2.**

In Figure 2, $G$ is well constrained. $I_1 = \{y_1, y_2, x_1, x_2\}$ is the unique irreducible down-subgraph of $G$. $I_2 = \{y_3, y_4, x_3, x_4\}$ is the unique down-irreducible subgraph of $G - I_1$. $G - I_1 - I_2 = \{y_5, y_6, y_7, x_5, x_6, x_7\}$ is irreducible.

As proved in the sequel, bipartite graphs associated with over-constrained systems admit also such a decomposition, but it is not unique.

This paper gives an efficient method to obtain such decompositions. Its cost is bounded by the time of the search of a maximum matching.

These two decompositions are interesting in two ways.

First, programming by constraints is still programming, and so there are always bugs ! One can even think that debugging constraints programs is more difficult than debugging imperative programs. Then the decomposition in well, over- and under- constrained parts is a first and important help. Moreover, when the system is well constrained, the decomposition into irreducible parts enables the user to follow the resolution process step by step, ie irreducible after irreducible : the decomposition gives the trace of running of the resolution process. This is a very important feature : remember that classical numerical methods cannot explain their way to the user.

On the other hand, for well constrained systems, the decomposition into irreducible ones greatly speeds up the resolution. Numerical methods of resolution needs times at least cubic in the size of the system, and symbolic methods needs exponential time, so any method which reduces the system into smaller ones is interesting. Maybe the system is irreducible, but anyway the time needed by this decomposition is negligible compared to the time of the brute force approach.

For over-constrained systems, the graph is not sufficient enough to decide wether the equations are redundant or contradictory. A possible way to solve such a system is to find any well constrained subsystem, to solve it (this system can also be reduced into irreducible components) and to verify if the remaining equations are satisfied or not.

For under-constrained systems, we do not have enough information to set all the unknowns. A possible way is to consider some of them as parameters. We can choose them to obtain a well constrained system, and then we can apply the previous method. Howewer, there is a problem in this choice : this will be detailed later.

Geometric modeling by constraints has received great attention. Barford has given a necessary and sufficient condition for a system to be well constrained, in terms of maximum network flows [Barford87]. Serrano in [Serrano91] has asked for the decomposition into subsystems, but he did not detail a polynomial algorithm. Owen in [Owen91] also used graphs and decomposition, but in a very different way. Murota in [Murota87] also used the Dulmage - Mendelsohn decomposition, but in a more sophisticated modelization : he differentiates unknowns and parameters and he uses a different bipartite graph. The Dulmage - Mendelsohn decomposition was sometimes used to solve large linear systems of equations, but there exist more specific and efficient methods in the linear case, see [Duff77] and [Poth-Fan90].

The results presented here are well known in graph theory but they seem to be ignored in the solid modeling community. This paper intends to be a comprehensive introduction to this theory, and to show its interest in practical resolution of systems in solid modeling.

The paper is organised as follows. In the second section we give some details on the relation between the algebraic system and the graph which modelises it. In the third section we analyze the mathematical structure of bipartite graph, we present the Dulmage-Mendelsohn and König decomposition and explain how to obtain them. In the fourth section we give the algorithms which calculate these decompositions, they are linear in space and their time is bounded by the cost of the search of a maximum matching in a bipartite graph, we give also three examples of constraint systems. We end the paper by some open problems.

## 2. ALGEBRAIC SYSTEM AND GRAPH

In this paragraph, we investigate some limitations of our modelization based on the fact that the knowledge of the structural properties of the associated bipartite graph is not sufficient enough to completely solve the system. Let us mention some of them.

Equations of an over-constrained system are either redundant or contradictory, but the graph cannot distinguish between these cases.

Generally, an under-constrained graph is associated with a system which has an infinite and not numerable set of solutions. For instance, the system with one equation and two unknowns : $\{ x_1 + x_2 = 0 \}$. However, some under-constrained systems have a finite (or empty) set of real solutions, for example the system with one equation and two unknowns : $\{ x_1^2 + x_2^2 = 0 \}$. The graph gives no help to distinguish between these cases.

"Structurally", a well constrained graph is associated with a system which has a finite number of solutions. However, in some "accidental" (by opposition to "structural") cases, the jacobian $|\partial f_i / \partial x_j|$ can be null for all $x_j$, and the graph cannot detect such a case.

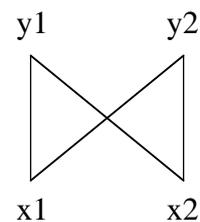

**Figure 3.**

For example, the graph of Figure 3 is associated with the well constrained system $\{x_1 + x_2 = 1, x_1 + 2x_2 = 3\}$ which has a finite set of solutions, with the singular system $\{ x_1 + x_2 = 1, 2x_1 + 2x_2 = 3\}$ which has no solutions, and with the singular system $\{ x_1 + x_2 = 1, 2x_1 + 2x_2 = 2\}$ which has an infinite and not numerable set of solutions. In the last two cases, the jacobian is null for all $x_j$, due to the fact that the coefficients satisfy a "parasite" condition : if you slightly modify some of the coefficients, without changing the associated bipartite graph, the jacobian does not vanish. In these cases, the nullity of the jacobian has nothing to do with the structure of the graph.

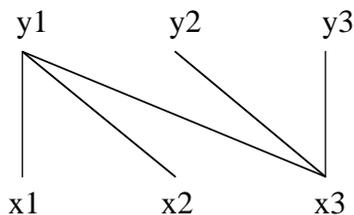

**Figure 4.**

On the contrary, the graph in Figure 4 is not well constrained (there is no perfect matching), so we can assert that the jacobian of all associated systems is "structurally" null, and no matter the values of the coefficients. In fact, the rank of the jacobian is always smaller or equal to the cardinality of a maximum matching of the associated bipartite graph. In general, these two values are equal. When they are different, it is due to "accidental" reasons, as said in the previous paragraph. See [Murota87].

In a more mathematical way, when all coefficients of the system are algebraically independent, then they cannot satisfy any parasite equation and the rank of the jacobian is equal to the cardinality of a maximum matching, see [Murota87].

The following concentrates on the structural properties of systems of equations and so ignores these "accidental" cases.

## 3. STRUCTURAL PROPERTIES OF BIPARTITE GRAPHS

Let us now present the fundamental results of Dulmage & Mendelsohn about the decomposition of bipartite graphs, see [Dulmage-Mendelsohn58], [Dulmage-Mendelsohn59], [Dulmage-Mendelsohn62], [Dulmage-Mendelsohn63]. The decomposition algorithm will be described in the next paragraph.

First we need to recall some definitions and to give some notations.
Let $G = (V, E)$ be a bipartite graph with edges $E$ and with vertices $V$. Then $V = Y \cup X$ and $Y \cap X = \emptyset$. $Y$ is the set of equations, $X$ is the set of unknowns.

A matching $M$ of $G$ is any subset of $E$ such that any two distinct edges in $M$ do not have a common vertex. $M$ is a maximum matching iff it is maximal in cardinality. A vertex is saturated, or covered, by $M$ iff it is a vertex of one edge in $M$. A matching saturating all vertices of $G$ is called perfect. We use all along the paper the classical notions of graph theory, see [Berge83].

Let us just recall the definition of connected components and of *strongly* connected components.

A graph $G = (V, E)$ is said to be connected iff for any pair $x$ and $y$ of vertices, there exists a *non directed path* joining $x$ and $y$ in $G$. The connected components of $G$ are the maximal connected subgraphs of $G$. They partition $V$ and $E$.

A *directed* graph $G = (V, E)$ is said to be *strongly* connected iff for any pair $x$ and $y$ of vertices, there exist a *directed path* from $x$ to $y$ and a *directed path* from $y$ to $x$. The *strongly* connected components of $G$ are the maximal *strongly* connected subgraphs of $G$. They partition $V$.

In a directed graph $G$, a vertex $s$ is said to be a source iff $G$ does not contain any arc $vs$ and $s$ is said to be a sink iff $G$ does not contain any arc $sv$.

### 3.1. Dulmage - Mendelsohn decomposition

As mentionned in the introduction, any bipartite graph can be canonically partitionned in three parts. The following theorem due to Dulmage & Mendelsohn describes this structure.

**Theorem 1.** Let $G = (V, E)$ be any bipartite graph $G$. Then $V$ can be partitionned into three sets : $D, A, C$ where $D$ is the set of all vertices in $G$ which are not covered by at least one maximum matching, $A$ is the set of all vertices in $V - D$ adjacent to at least one vertex in $D$ and finally $C$ is $V - A - D$. These subsets are unique and yield a unique decomposition of $G$ into three subgraphs $G_1, G_2, G_3$ defined by $G_1 = (C_1, C_2, E_1)$ where $C_1 = C \cap Y$, $C_2 = C \cap X$, and $E_1$ are induced edges of $G_1$, $G_2 = (D_1, A_2, E_2)$ where $D_1 = D \cap Y$, $A_2 = A \cap X$, and $E_2$ are induced edges of $G_2$, $G_3 = (A_1, D_2, E_3)$ where $A_1 = A \cap Y$, $D_2 = D \cap X$, and $E_3$ are induced edges of $G_3$.

**Proof :** see [Lovasz-Plummer86].

An example of such a decomposition, called DM decomposition for short, is shown in Figure 5. At this point, remark that $G_1$ or $G_2$ or $G_3$ can be empty. On the other hand, the DM decomposition has been extended to general graphs by Gallaï and Edmonds, see [Lovasz-Plummer86].

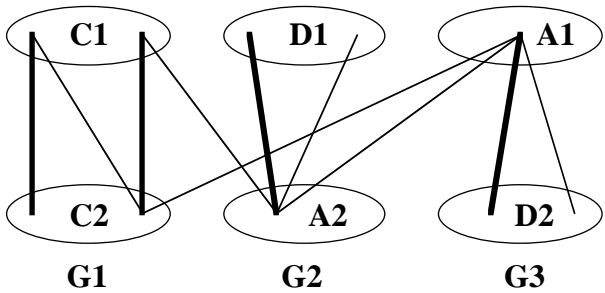

**Figure 5.**

This decomposition has the following properties :

• There is no edge between $D_1$ and $C_2$, between $D_2$ and $C_1$, between $D_1$ and $D_2$.

• $G_1$ has a perfect matching, so $|C_1| = |C_2|$.

• Every maximum matching of $G$ consists of a perfect matching of $G_1$, a matching of $A_1$ into $D_2$ (all vertices of $A_1$ are covered, at least one vertex of $D_2$ is not), and a matching of $A_2$ into $D_1$ (all vertices of $A_2$ are covered, at least one vertex of $D_1$ is not). So $|M| = |C_1| + |A_1| + |A_2|$ and $|D_1| > |A_2|$, $|A_1| < |D_2|$.

• Edges between $C_1$ and $A_2$, between $C_2$ and $A_1$, between $A_1$ and $A_2$ never belong to a maximum matching.

$G_1$ corresponds to the well constrained part of the system, $G_2$ to the over-constrained part, and $G_3$ to the under-constrained part.

### 3.2. Bipartite graphs with a perfect matching

Let us now consider a bipartite graph $G = (V, E)$ with a perfect matching : the associated system is structurally well constrained. $G$ is defined to be irreducible iff for every proper subset $Z$ of $Y$, $|\Gamma(Z)| > |Z|$. Equivalently, $G$ is irreducible iff for each edge $e$ of $G$ there always exists a perfect matching containing $e$, see [Berge83]. In other words, with our model, an irreducible graph corresponds to a well constrained system whose all proper subsystems are under-constrained.

Another result due to König, and Dulmage & Mendelsohn gives the unique and canonical decomposition of any bipartite graph with a perfect matching into irreducible subgraphs :

**Theorem 2.** Let $H$ be the graph obtained from $G$ after deleting all edges which never belong to a perfect matching of $G$. Then the connected components $H_1, H_2, \ldots H_k$ of $H$ are irreducible.

**Proof.** see [Lovasz-Plummer86].

An example is shown in Figure 6.

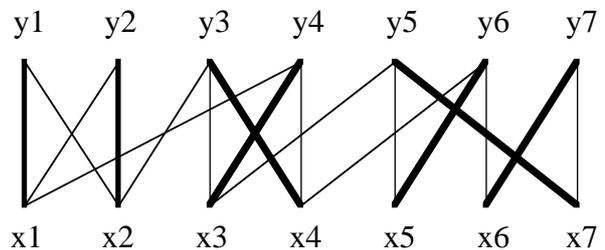

A perfect matching of a bipartite graph $G$.

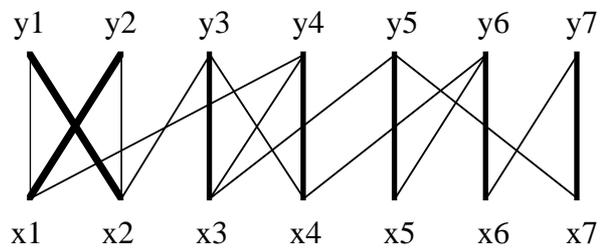

Another perfect matching of G.

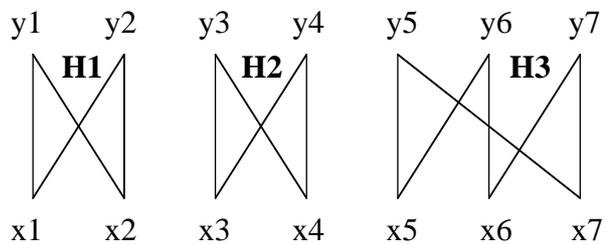

The graph $H$.

**Figure 6.**

The question is now to have an efficient way to find this decomposition. The following property gives an answer to this question.

**Property 1.** Consider *any* perfect matching $M$ of $G$. Let $G'$ be the directed graph obtained from $G$ by replacing each edge $xy$ in $M$ by the two arcs $xy$ and $yx$, and by orienting all the other edges from $Y$ to $X$. Then the *strongly* connected components of $G'$ are exactly the connected components of $H : H_1, H_2, ... H_k$.

**Proof.** It is sufficient to note that we have : $G$ is irreducible $\Leftrightarrow$ $G'$ is *strongly* connected. This equivalence can be easily proved. So the *strongly* connected components of $G''$ are irreducible, they necessarily correspond to the unique $H_1, H_2, ... H_k$ of Theorem 2.

In fact this construction yields more information. Let $R$ be the *directed* graph obtained from $G$ by contracting in one vertex each *strongly* connected component of $G'$ *(or H)*. For instance the previous graph leads to the graph $R$ shown in Figure 7.

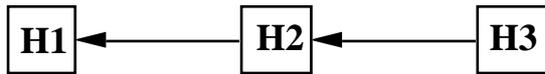

**Figure 7.**

It is well known that $R$ is acyclic, so $R$ induces a partial order on $H_1, H_2, ... H_k$. For our needs, this means that if $R$ has an arc from $H_i$ to $H_j$, then subsystem $H_i$ uses some unknown(s) of subsystem $H_j$; thus $H_j$ must be solved before $H_i$.

### 3.3. General case

Let us now consider any bipartite graph $G$ and let $M$ be any maximum matching of $G$. We define $G'$ to be the directed graph obtained from $G$ by replacing each edge $xy$ in $M$ by two arcs $xy$ and $yx$, and by orienting all other edges from $Y$ to $X$. The *strongly* connected components of $G'$ are included either in $G_1$, or in $G_2$, or in $G_3$. Moreover, if $Y$ contains non saturated vertices, then they are sources of $G_2$, thus $G_2$ is not empty. Symmetrically if $X$ contains non saturated vertices, then they are sinks of $G_3$, which is therefore not empty. These properties result directly from the definition of $G'$, from Theorem 1 and its consequences.

Then the structure obtained on $G'$ has necessarily the form given in Figure 8.

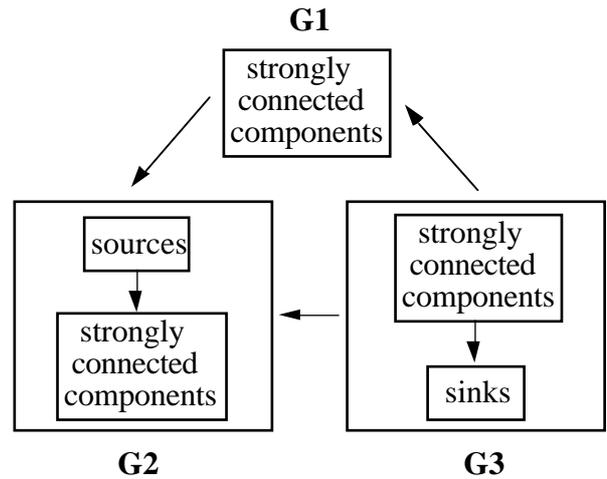

**Figure 8.**

The property 1 ensures a unique decomposition for $G_1$, but the decompositions of $G_2$ and $G_3$ depends on the maximum matching $M$ chosen. Anyway we have the following property.

**Property 2.**

$G_2 = \{ z \mid \exists$ path $y,..., z$ in $G'$ such that $y$ is a source of $G'$ $\}$

$G_3 = \{ t \mid \exists$ path $t,..., x$ in $G'$ such that $x$ is a sink of $G'$ $\}$

**Proof.** It is an immediate consequence of the structure of $G'$.

Similarly to the case of graphs with a perfect matching, the directed graph R, obtained from $G'$ by contracting in one vertex each *strongly* connected component, is acyclic, this induces a partial order between *strongly* connected components. Thus any compatible total order gives an order of resolution of subsystems associated with *strongly* connected components of $G_1$ and $G_2$ with cardinality strictly greater than 1. This is a consequence of the fact that $G_1 \cup G_2$ always contains an irreducible down-subgraph. On the other hand, as $G_3$ cannot contain an irreducible down-subgraph, there is no way to solve it, except by seting unknowns corresponding to its non saturated vertices.

It is important to mention that the order of resolution on $G_2$ is fundamentally dependent on the choice of the maximum matching. In fact, $G_2$ contains several irreducible down-subgraphs, and the best method is to solve the smallest in cardinality. However, efficiently finding the smallest irreducible down-subgraph of $G_2$ is an open problem. Another strategy is to choose any subset $S$ of $D_1$ with cardinality $|A_2|$;

the down-subgraph generated by *S* and denoted by *T* has a perfect matching under the hypothesis that $G_2$ has only one connected component. Otherwise we have to work on each connected component of $G_2$. Then we apply the decomposition given in 3.2 to this subgraph *T*.

A similar problem arises with $G_3$. We can set any subset of unknowns in $D_2$ with cardinality $|D_2| - |A_1|$, for instance the non saturated vertices of $D_2$. The resulting subgraph *T* has a perfect matching under the hypothesis that $G_3$ has only one connected component. Otherwise we have to work on each connected component. Then we apply the decomposition given in 3.2 to this subgraph *T*. A problem here is to find a good subset *S*, that means one that will yield a subgraph *T* with the smallest irreducible down-subgraph. This seems to be a difficult problem.

For instance, consider the graph *G* shown in Figure 9 and a maximum matching $M = \{ y_1x_1, y_2x_2, y_4x_3, y_6x_4, y_7x_6 \}$ :

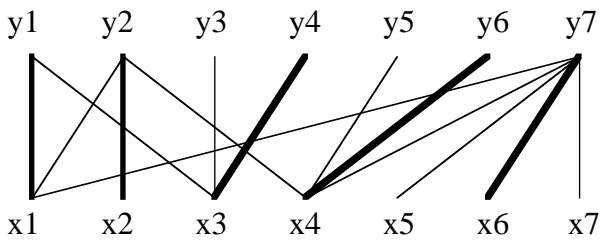

**Figure 9.**

$y_3$ and $y_5$ are the non saturated vertices of *Y*, and so the sources of $G_2$. By property 2, we get $G_2 = \{y_3, x_3, y_4, y_5, x_4, y_6\}$, shown in Figure 10. Note that $G_2$ has two connected components $\{y_3, x_3, y_4\}$ and $\{y_5, x_4, y_6\}$.

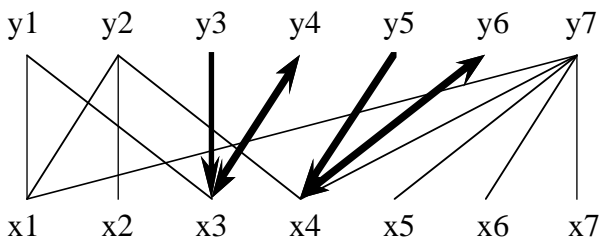

**Figure 10.**

$x_5$ and $x_7$ are the non saturated vertices of *X*, and so the sinks of $G_3$. By property 2, we get $G_3 = \{y_7, x_5, x_6, x_7\}$, shown in Figure 11.

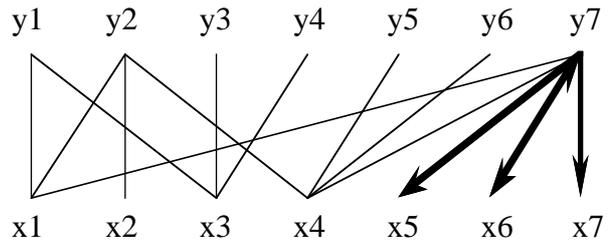

**Figure 11.**

Finally we get $G_1 = G - G_2 - G_3 = \{x_1, x_2, y_1, y_2\}$.

## 4. ALGORITHMS

Let $G = (V, E)$ be a bipartite graph associated with a system of equations. Let us take the notations $n = |V|$, $m = |E|$. We now give the algorithms to obtain the previously presented decompositions. Their proofs are direct consequences of the properties described in section 3 :

### 4.1. DM - Decomposition

The subgraphs $G_1$, $G_2$, $G_3$ of *G* can be obtained by the following algorithm :

1. Find a maximum matching *M* of *G*.

2. Build the directed graph *G'* from *G* by replacing each edge *xy* in *M* by two arcs *xy* and *yx*, and by orienting all other edges from *Y* to *X*.

3. $G_2$ is the set of all descendants of sources of *G'*.

4. Symmetrically, $G_3$ is the set of all ancestors of sinks of *G'*.

5. Finally, $G_1$ is $G - G_2 - G_3$.

The steps 2, 3, 4 and 5 can be computed in $O(n+m)$, and the whole algorithm runs in $O(m n^{1/2})$ using Hopcroft & Karp's algorithm [Hopcroft-Karp73] for step 1.

The computation of the connected components of $G_1$, $G_2$, $G_3$ can be done by using depth first or breadth first search in linear time.

### 4.2. Decomposition into irreducible parts

#### 4.2.1. Well constrained systems

Suppose $G$ is well constrained, $G = G_1$ and $G_2 = G_3 = \emptyset$. The following algorithm gives the unique decomposition of $G$ into its irreducible components and an order of resolution between them (see Theorem 2).

1. Find a maximum matching $M$ of $G$ (actually, $M$ is a perfect matching).

2. Built the directed graph $G'$ from $G$ by replacing each edge $xy$ in $M$ by two arcs $xy$ and $yx$, and by orienting all other edges from $Y$ to $X$.

3. Compute the *strongly* connected components of $G'$. Each of these *strongly* connected components is irreducible.

4. To compute the dependencies between these irreducible subgraphs, build the reduced graph $R$ from $G'$ by contracting each *strongly* connected component in a vertex. Each arc of $R$, say from $s_1$ to $s_2$, means : solve subsystem $s_2$ before $s_1$. A compatible total order between subsystems can be obtained by any topological sorting of $R$.

Steps 2, 3 and 4 can be computed in $O(n+m)$. Step 3 and 4 can be executed using Tarjan's algorithm [Tarjan72]. At this point, we can note that the order of obtention of the *strongly* connected components (ie the irreducible subsystems) is a compatible order of resolution, it is a byproduct of Tarjan's algorithm. Thus the step 4 can be suppressed if we do not want the partial order induced by $R$. Running time of the whole algorithm including step 1 is bounded by the cost of the search of a maximum matching. Of course, if a maximum matching is already known, the algorithm is linear.

The decomposition is independent of the maximum matching $M$.

### 4.2.2. Well and over-constrained systems

The previous method can be applied to $G_1 \cup G_2$. However, the maximum matching $M$ is not perfect, and remember that the decomposition of $G_2$ depends on the maximum matching $M$.

The method corresponds to reject the non saturated vertices of $G_2$. Thus we obtain a well constrained system which can be completely solved. At the end, we have to verify that the discarded equations are satisfied by the found solutions.

### 4.3. Experiments

We present here three illustratives examples of real systems of constraints and give for each the corresponding decomposition.

A "dimensioning scheme" is shown in figure 12. The points $A$ and $B$ are initially fixed. The labelled edges corresponds to the distance constraints (quadratic equations). $ß_1$ and $ß_2$ are the arguments of the angle constraints.

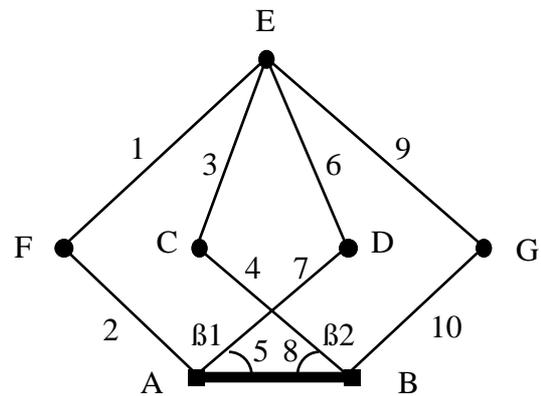

**Figure 12.**

The system of equations corresponding to this scheme is well constrained so $G = G_1$ and $G_2 = G_3 = \emptyset$. The perfect matching $\{eq_1 x_C, eq_2 y_C, eq_3 x_D, eq_4 y_D, eq_5 x_E, eq_6 y_E, eq_7 x_F, eq_8 y_F, eq_9 x_G, eq_{10} y_G\}$ and the decomposition of $G$ into irreducible components is shown in figure 13. The order of resolution of these irreducible parts is the following : $G_{11}, G_{12}, G_{13}, G_{14}$ and $G_{15}$. We first compute the coordinates of the point $C$ using the two first equations, then the coordinates of points $D, E, F$ and $G$ in this order.

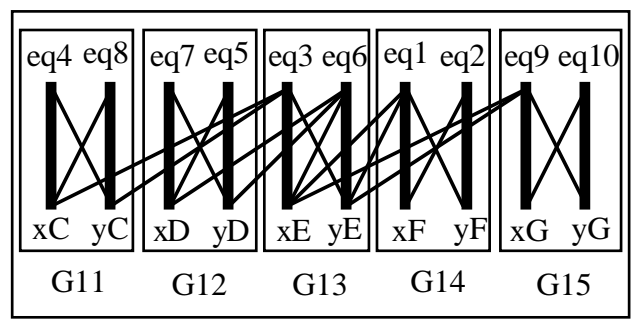

**Figure 13.**

In practice, to solve algebraic systems, we use a variant of Krawczyk - Moore algorithm : a bisection method that use interval Newton iterations (see [Kearfott87], [Moore-Qi82] or [Snyder92]). This method finds all solutions in a given domain. The resolution time of figure 12 is divided by 20 when using this decomposition instead of methods that solve simultaneously the set of equations. The gain can be more important for big reducible systems. For irreducible systems, the decomposition can not speed up the resolution but the decomposition time is anyway negligible compared to the resolution time.

Both under and over constrained systems causes numerical difficulties if solved by classical methods. The decomposition we used here allow us detecting under and over constrained geometries of a scheme. An example of an under-constrained dimensioning scheme is illustrated in figure 14. The point $E$ is free to move along the circle centered in $C$ with the distance between points $C$ and $E$ as ray. As shown above, the points $A$ and $B$ are fixed.

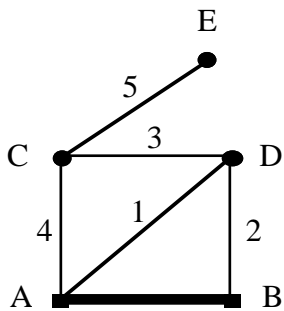

**Figure 14.**

The decomposition of the associated graph is shown in figure 15.

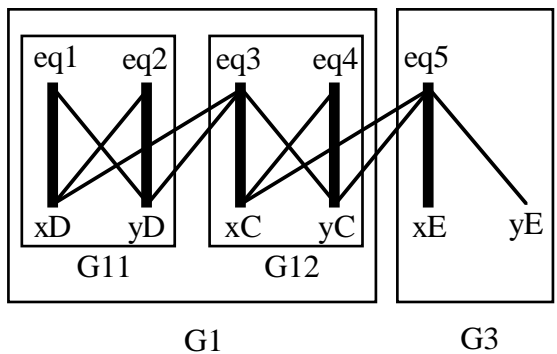

**Figure 15.**

$G = G_1 + G_3, G_2 = \emptyset$.

The sub-graph $G_3$ is not empty so we have an under-constrained geometry (here the point $E$).

An example of an over-constrained dimensioning scheme is illustrated in figure 16.

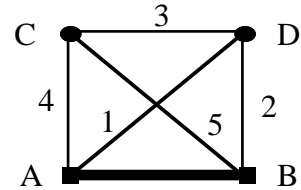

**Figure 16.**

The decomposition of the associated graph is shown in figure 17.

$G = G_2, G_1 = G_3 = \emptyset$.

The sub-graph $G_2$ is not empty so we have conflicting constraints. A strategy is to solve the system of equations after removing (eq5) (the non saturated equation vertex) and then test if the found solution verifies (eq5).

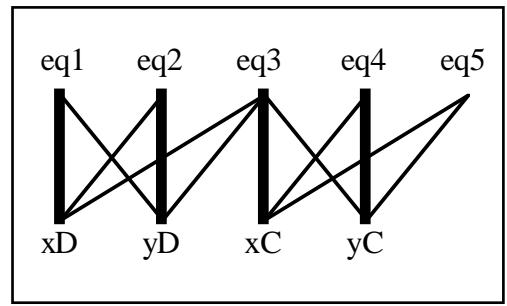

**Figure 17.**

This work was implemented on a SUN workstation using LeLisp language.

## 5. CONCLUSION

The methods presented in this paper can be used to gain some knowledge on the combinatorial structure of the systems of equations, and so to debug them. On the other hand, they significantly speed up the resolution and enable the user to follow the resolution process.

However the problem is not completely solved for over- and under-constrained systems. In the first case, efficiently finding the smallest irreducible down-subgraph of the associated graph and, in the second case, finding the best subset of unknowns to set, seems to be interesting problems of graph theory.

Previous algorithms build decompositions from scratch, but for interactive use, the possibility to incrementally modify the systems of constraints is useful. In this context, we have to incrementally and efficiently update the decompositions after each modification : insertion or suppression of a vertex or an edge. This is another open question.